\documentclass[10pt,letterpaper]{article}
\usepackage[top=0.85in,left=2.75in,footskip=0.75in]{geometry}

\usepackage{amsmath,amssymb}

\usepackage{changepage}

\usepackage{textcomp,marvosym}

\usepackage{nameref,hyperref}

\usepackage[right]{lineno}

\usepackage[nopatch=eqnum]{microtype}
\DisableLigatures[f]{encoding = *, family = * }

\usepackage[table]{xcolor}

\usepackage{array}
\usepackage{float}
\usepackage{subfig}
\usepackage[numbers,sort&compress]{natbib}
\newcolumntype{+}{!{\vrule width 2pt}}

\newlength\savedwidth

\raggedright
\usepackage[aboveskip=1pt,labelfont=bf,labelsep=period,justification=raggedright,singlelinecheck=off]{caption}

\makeatletter
\renewcommand{\@biblabel}[1]{\quad#1.}
\makeatother

\usepackage{lastpage,fancyhdr,graphicx}
\usepackage{epstopdf}
\fancyheadoffset[L]{2.25in}
\fancyfootoffset[L]{2.25in}
\begin{document}
\vspace*{0.2in}

\begin{flushleft}
{\Large
\textbf\newline{Improving Lung Cancer Diagnosis and Survival Prediction with Deep Learning and CT Imaging} % Please use "sentence case" for title and headings (capitalize only the first word in a title (or heading), the first word in a subtitle (or subheading), and any proper nouns).
}
\newline
\\
Xiawei Wang \textsuperscript{1},
James Sharpnack\textsuperscript{2,},
Thomas C.M. Lee\textsuperscript{1,2,*},
\\
\bigskip
\textbf{1} Graduate Group in Biostatistics, University of California Davis, Davis, CA, USA\\
\textbf{2} Department of Statistics, University of California Davis, Davis, CA, USA
\\
\bigskip

% Insert additional author notes using the symbols described below. Insert symbol callouts after author names as necessary.
% 
% Remove or comment out the author notes below if they aren't used.
%
% Primary Equal Contribution Note
%\Yinyang These authors contributed equally to this work.

% Additional Equal Contribution Note
% Also use this double-dagger symbol for special authorship notes, such as senior authorship.
%\ddag These authors also contributed equally to this work.

%\textcurrency Current Address: 1045 La Avenida St, Mountain View, CA 94043\\
%\textcurrency Current Address:
%399 Crocker Lane,
%University of California Davis,
%One Shields Avenue
%Davis, CA, USA %$ Dept/Program/Center, Institution Name, City, State, Country % change symbol to "\textcurrency a" if more than one current address note
% \textcurrency b Insert second current address 
% \textcurrency c Insert third current address

% Group/Consortium Author Note
%\textpilcrow Membership list can be found in the Acknowledgments section.

%*Corresponding author e-mail: tcmlee@ucdavis.edu

\end{flushleft}
\section*{Abstract}
Lung cancer is a major cause of cancer-related deaths, and early diagnosis and treatment are crucial for improving patients' survival outcomes. In this paper, we propose to employ convolutional neural networks to model the non-linear relationship between the risk of lung cancer and the lungs' morphology revealed in the CT images. We apply a mini-batched loss that extends the Cox proportional hazards model to handle the non-convexity induced by neural networks, which also enables the training of large data sets. Additionally, we propose to combine mini-batched loss and binary cross-entropy to predict both lung cancer occurrence and the risk of mortality. Simulation results demonstrate the effectiveness of both the mini-batched loss with and without the censoring mechanism, as well as its combination with binary cross-entropy. We evaluate our approach on the National Lung Screening Trial data set with several 3D convolutional neural network architectures, achieving high AUC and C-index scores for lung cancer classification and survival prediction. These results, obtained from simulations and real data experiments, highlight the potential of our approach to improving the diagnosis and treatment of lung cancer.

% Please keep the Author Summary between 150 and 200 words
% Use first person. PLOS ONE authors please skip this step. 
% Author Summary not valid for PLOS ONE submissions.   
%\section{Author summary}

%\linenumbers

% Use "Eq" instead of "Equation" for equation citations

\section{Introduction}

Lung cancer is one of the most common causes of cancer-related deaths worldwide. Early diagnosis and treatment are crucial for improving patients' survival rates \cite{alberg2003epidemiology,spiro2005one}. Survival analysis, a branch of statistics that has been widely used in public health research, provides valuable insights into the impact of different conditions on the survival time of patients; e.g., \cite{ishaq2021improving,lee2019dynamic}. In the context of lung cancer, early detection through screening methods helps identify the tumor in its early stage and applying survival analysis to lung cancer patients can aid in early detection and ultimately improve patients' survival outcomes. Meanwhile, in recent years, computer-aided diagnosis has gained significant attention, particularly in medical image data analysis \cite{chen2021multi,du2022parameter,Li2018,mielke2009regionally,du2024autogfi}. Deep learning techniques have been increasingly applied to analyze various kinds of medical images due to their effectiveness, for example, \cite{hou2016patch,gao2019distanced,wang2019lung,ardila2019end,liu2020no,zhang2020image,10526478}.

%\xwang{This paragraph will be replaced by the blue text below, I kept it for comparison of the citations, [10-16] are the newly cited papers. }

%\xwang{Statistical methods play a crucial role in enhancing our understanding of diseases, contributing to early diagnosis and improving treatment outcomes for patients \cite{alberg2003epidemiology, spiro2005one, mielke2009regionally}. Survival analysis, a branch of statistics that has been widely used in public health research, provides valuable insights into the impact of different conditions on the survival time of patients, as demonstrated in studies such as \cite{ishaq2021improving, lee2019dynamic}. In recent years, computer-aided diagnosis has gained significant attention, particularly in medical image data analysis. Deep learning techniques have been increasingly applied to analyze various types of medical images due to their effectiveness, for example, \cite{zhang2020image,bakas2018identifying,zhong2023multi,du2022parameter,chen2021multi}.Lung cancer stands as one of the leading causes of cancer-related deaths globally, impacting both men and women. The integration of survival analysis into the assessment of lung cancer patients holds the potential to contribute to early detection, ultimately leading to improved survival outcomes. Existing work predominantly focused on lung cancer patients' medical images alone, \cite{hou2016patch, gao2019distanced, wang2019lung, ardila2019end, liu2020no}.} 

Despite the promising results obtained by using these techniques, the accessibility of high-quality medical images poses a challenge in applying these techniques. For example, Hou \textit{et al.} \cite{hou2016patch} required whole slide tissue images obtained from invasive procedures, Gao \textit{et al.} \cite{gao2019distanced} required multiple longitudinal CT images captured over time, and Wang \textit{et al.} \cite{wang2019lung} required both demographic information and chest CT images. 

In addition, most of these studies focused on patients already diagnosed, neglecting those who may be prospective candidates undergoing regular CT screening for early detection.  Furthermore, there are few works that have utilized survival analysis, which limits the statistical efficiency of these methods. Considering the significant impact of early detection on patients' survival chances \cite{blandin2017progress}, there is an urgent need to develop a new approach that can enhance both the early detection and survival prediction for individuals currently diagnosed and those potentially at risk of lung cancer, while considering the accessibility of the medical image data. 

This paper aims to utilize deep learning techniques to analyze the potential lung cancer patients' survival hazards only based on their most recent CT images. Inspired by DeepSurv \cite{katzman2018deepsurv}, which uses demographic information, and DeepConvSurv \cite{zhu2016deep}, which uses 2D pathological images, we adopt 3D convolutional neural networks (CNNs) to model the non-linear relationship between the risk of lung cancer and the lungs' morphology revealed in CT images. A mini-batched loss involving time-to-event and censoring status is applied for handling the non-convexity caused by the neural networks and allowing the training of large data sets at the same time. In addition, we propose to apply the combination of binary cross-entropy and the mini-batched loss to simultaneously predict whether a potential patient has lung cancer and the risk of dying from it. The promising empirical properties of the proposed method are illustrated by simulation experiments and the application to the National Lung Screening Trial (NLST) dataset \cite{national2011national}. 

Our approach has several distinct features: (i) it relates patients' survival with 3D medical image classification; (ii) it considers both existing and potential patients, which helps in the early detection of the disease; and (iii) it requires only one raw CT scan, eliminating the need for additional clinical or longitudinal data or human pathologists' annotation, which makes our approach easy to implement and more accessible than methods that require extensive data collection.

The rest of this paper is organized as follows: Section~\ref{sec:background} introduces related works in computer-aided diagnosis and basic knowledge about survival data and the Cox proportional hazards model. Section~\ref{sec:mthd} derives the mini-batched loss function of the extended Cox model and introduces the idea of the two-task method and corresponding metrics. Section~\ref{sec:simu} presents the simulation study of the mini-batched loss based on the MNIST dataset and the simulation of the two-task method based on the Nodule-CIFAR dataset. Section~\ref{sec:exp} presents the real data experiment with the two-task method, which includes CT images from potential lung cancer patients.

\section{Background}
\label{sec:background}
\subsection{Related Work}
The Cox proportional hazards model \cite{cox1972regression} was first proposed to explore the relationship between the survival chance of a patient and a group of explanatory variables through the concept of hazard rate, see Eq~\ref{cox}. Later, Breslow \cite{breslow1972discussion} and Cox \cite{cox1975partial} discussed the estimation of model parameters, particularly for the baseline hazard function. Despite it being proposed more than 50 years ago, the Cox model continues to be one of the most widely used models in medical research for investigating patients' survival chances.

The use of medical images to aid the diagnosis and treatment of diseases has become increasingly popular. 
%Medical images have become an important tool for the diagnosis and treatment of different diseases. 
Much research has been conducted on the use of deep learning techniques to analyze medical images as a computer-aided diagnosis. For example, Hou \textit{et al.}\cite{hou2016patch} studied the feature of whole slide tissue image patches with a CNN. Wang \textit{et al.} \cite{wang2019lung} detected lung cancer with CT images and clinical demographics. Ardila \textit{et al.} \cite{ardila2019end} proposed a CNN-based method to predict lung cancer risk. Gao \textit{et al.} \cite{gao2019distanced} performed research in detecting lung cancer with long short-term models. Liu \textit{et al.} \cite{liu2020no} studied detecting nodules from CT images for lung cancer with adversarial attacks. However, some of these images or data may not be readily available or collected. These methods required whole slide tissue images from an invasive procedure \cite{hou2016patch}, or longitudinal medical images captured over time \cite{gao2019distanced,ardila2019end}, or demographic information in addition to medical images \cite{wang2019lung}. For more details, refer to \cite{cao2020comprehensive,singh20203d} for a comprehensive review of deep learning techniques applied to medical images. 

While these imaging methods have produced excellent results for the tasks that they were designed for, they did not establish a correlation with patients' survival.  
%While these methods are advanced in analyzing medical images, they did not establish a correlation with patients' survival. 
Katzman \textit{et al.} \cite{katzman2018deepsurv}, for the first time, developed the DeepSurv model to study the non-linear relationship between survival hazards and clinical features. It replaced the linear part $\boldsymbol{{\beta}}^\intercal\boldsymbol{x}$  in the Cox proportional hazards model (\ref{cox}) with multi-layer perceptrons $f(\boldsymbol{x})$. However, this model has a limitation in that it can only process clinical information. To address this limitation, DeepConvSurv was then proposed by Zhu \textit{et al.} to predict patients' survival directly from the 2D region of interests (ROI) of pathological images, using CNNs for $f(\boldsymbol{x})$. 

In this paper, we aim to expand previous research by developing a model that classifies lung cancer occurrence from potential lung cancer patients with only one 3D CT scan and further predicts the patient's relative hazards of dying from lung cancer. Our approach integrates 3D CNNs, binary classification, and the Cox proportional hazards model. By combining these techniques, we aim to establish a direct correlation between potential patients' 3D medical images and patients' survival, which could have significant implications for early lung cancer diagnosis. 

\subsection{Survival Data}
Survival analysis typically considers time-to-event data. Let $T^* = \min(T, C)$ be the observed time, where $T$ denotes the event time and $C$ denotes the censored time. Here, $T$ is the time from the beginning of the observation to an event, usually death, disease occurrence, or other experience of interest, which can be unobserved if censoring occurs first. The censored time $C$ is the time after which nothing is observed about the object. In addition to observing $T^*$, we also have the event indicator: $\delta_i = 1_{\{T_i\le C_i\}}$ that tells us if the $i$-th observation $T_i$ is censored or not. In our study,  $T^*$ is the observed time from the beginning of the study to either observed death or censoring. If death is observed, $T^* = T$ and $\delta = 1$, if censoring is observed, $T^* = C$ and $\delta = 0$.
The objective is to model the event distribution of $T$,  
\begin{equation*}
    F(t) = P(T\leq t ) = \int_{0}^{t}f(u)du,
\end{equation*}
where the density function $f(t)$ is  
\begin{equation*}
    f(t) = \lim_{\Delta t \to 0} \frac{P(t< T \le t + \Delta t)}{\Delta t}.
\end{equation*}
%\tlee{We usually don't put a number at the end of an equation if the equation won't be referenced at all.} \xwang{fixed}

In survival analysis, it is common to alternatively study the survival function $S(t)$, or the hazard function $\lambda(t)$, or the cumulative hazard function  $\Lambda(t)$, defined respectively as
\begin{equation*}
    S(t)  = P(T> t ) = \int_{t}^{\infty}f(u)du,
\end{equation*}
\begin{equation*}
    \lambda(t) =  \lim_{\Delta t \to 0 } \frac{P(t< T \le t + \Delta t|T> t)}{\Delta_t}, 
\end{equation*}
and
\begin{equation*}
    \Lambda(t) = \int_{0}^{t}\lambda(u)du. 
\end{equation*}
Their relationships can be expressed as 
\begin{equation*}
    \lambda(t) = \frac{f(t)}{S(t)}, 
\end{equation*}
and 
\begin{equation*}
S(t) = \exp{(-\Lambda(t)}),
\end{equation*} 
so it's equivalent to studying either of them. In this paper, we focus on the density function $f(t)$ and the corresponding likelihood function.

Given a set of right-censored samples $\{T^*_i,\delta_i\}_{i = 1}^n$, the likelihood function $L$ is:
\begin{equation*}
\begin{split}  
L  &= \prod_{i =1}^{n} f(T^*_i)^{\delta_i}S(T^*_i)^{1-\delta_i} \\
&= \prod_{i =1}^{n}\lambda(T^*_i)^{\delta_i}S(T^*_i),
\end{split}
\end{equation*} which can be further used for parameter estimation.

\subsection{Cox Proportional Hazards Model and DeepSurv}
The Cox proportional hazards model is one of the most used models for exploring the relationship between the hazards $\lambda(t|\boldsymbol{x})$ and the explanatory covariates $\boldsymbol{x}$. 
In particular, it assumes proportional hazards and linear contribution of the covariates to the log relative hazards function: 
\begin{equation}
\lambda(t|\boldsymbol{x}) = \lambda_0(t)\exp(\boldsymbol{{\beta}}^\intercal\boldsymbol{x}),
\label{cox}
\end{equation} 
where $t$ represents time, $\lambda_0(t)$ is the baseline hazard function (an infinite dimensional parameter), $\boldsymbol{x}$ is a set of covariates, and $\boldsymbol{\beta}$ is the corresponding coefficient that measures the effect of the covariates. 
%$\lambda(t|\boldsymbol{x})$ is the hazard function determined by baseline hazard $\lambda_0(t)$ and the relative hazards function $\exp(\boldsymbol{\beta}^\intercal\boldsymbol{x})$. 
Cox \cite{cox1972regression,cox1975partial} proposed to use the partial likelihood for estimating $\boldsymbol{\beta}$ with the advantage of circumventing $\lambda_0(t)$. Let $R(t) = \{i: T^*_i>t\}$ be the risk set at time $t$; i.e., the set of all individuals who are "at risk" for failure at time $t$. The partial likelihood is the product of the conditional probabilities of the observed individuals being chosen from the risk set to fail:
\begin{equation*}  L(\boldsymbol{\beta})_{\rm partial} = \prod_{i =1}^{n} \left[\frac{\exp(\boldsymbol{{\beta}}^\intercal\boldsymbol{x_i})}{\sum_{j\in R(T^*_i)}\exp(\boldsymbol{{\beta}}^\intercal\boldsymbol{x_j})}\right]^{\delta_i},
\end{equation*}
where $R(T^*_i)$ denotes the set of individuals that are ``at risk" for failure at time $T^*_i$ in the sample.

The estimate $\hat{\boldsymbol{\beta}}$ for $\boldsymbol{\beta}$ can be obtained by minimizing the averaged negative partial log-likelihood $\mathcal{L}(\boldsymbol{\beta})$, which is convex:
\begin{equation*}
 \mathcal{L}(\boldsymbol{\beta}) = -\frac{1}{n}\sum_{i=1}^{n} \delta_i \biggl[\boldsymbol{\beta}^\intercal \boldsymbol{x_i} -\log \sum_{j\in R(T^*_i)}\exp(\boldsymbol{{\beta}}^\intercal\boldsymbol{x_j})\biggr].   
\end{equation*} 
The cumulative baseline hazard function can be estimated with the Breslow estimator: 
\begin{equation*}
    \begin{split}
      \hat{\Lambda}_0(t; \boldsymbol{\beta}) &= \sum\limits_{j \notin R(t)}\Delta\hat{\Lambda}_0(T^*_j)\\
      &=  \sum\limits_{j \notin R(t)}\frac{\delta_j}{\sum\limits_{k\in \mathcal R(T^*_j)} \exp(\boldsymbol{\beta}^\intercal \boldsymbol{x_k})}.  
    \end{split}
\end{equation*}

The DeepSurv method can be seen as a non-linear version of the Cox model. It replaces the linear log relative hazards term $\boldsymbol{\beta}^\intercal\boldsymbol{x}$ in the Cox model with a non-linear multi-layer perceptron (MLP)  $f(\boldsymbol{x};\boldsymbol{\theta})$:
\begin{equation*}
\lambda(t|\boldsymbol{x}) = \lambda_0(t)\exp\bigl(f(\boldsymbol{x};\boldsymbol{\theta})\bigr),
\end{equation*} 
where $f(\boldsymbol{x};\boldsymbol{\theta})$ is a fully-connected MLP parameterized by $\boldsymbol{\theta}$.
%In this case, the loss function of DeepSurv is the averaged negative partial log-likelihood:
%\begin{equation*}
    %\mathcal{L}(\boldsymbol{\theta})= -\frac{1}{n} \sum\limits_{i=1}^{n} \delta_i\biggl[ f(\boldsymbol{x_i}; \boldsymbol{\theta}) - \log\sum\limits_{j\in  \mathcal R(T^*_i)} \exp{\bigl({f(\boldsymbol{x_j}; \boldsymbol{\theta})\bigr) }}\biggr].
%\end{equation*}

\section{Methodology}\label{sec:mthd}
\subsection{Extended Cox Model with Convolution Neural Network}
In this study, we modeled patients' hazard function of a certain disease based on 3D medical images. We cannot directly apply the DeepSurv or DeepConvSurv model because MLP or 2D CNN is deficient for 3D image data. Therefore, we extended the DeepSurv model by replacing MLP with a 3D convolution neural network $f(\boldsymbol{x};\boldsymbol{\Theta})$, which predicted the effects of a patient's morphological features $\boldsymbol{x}$ on their hazard rate and parameterized by the weights of the network $\Theta$:
\begin{equation*}
    \lambda(t|\boldsymbol{x}) = \lambda_0(t)\exp\bigl(f(\boldsymbol{x};\boldsymbol{\Theta})\bigr).
\end{equation*}

\subsection{Loss Function Derivation}
Let 
\begin{equation*}
\Lambda(t) = \Lambda_0(t)\exp\bigl(f(\boldsymbol{x};\boldsymbol{\Theta})\bigr)
\end{equation*}
and 
\begin{equation*}
    S(t) = \exp\Bigl(-\Lambda_0(t)\exp\bigl(f(\boldsymbol{x};\boldsymbol{\Theta})\bigr)\Bigr),
\end{equation*}
so the full likelihood function is
\begin{equation*} 
\begin{split}
L(\Lambda_0,\boldsymbol{\Theta})  = 
\prod_{i=1}^n \biggl\{&\Bigl[\lambda_0(T^*_i)\exp\bigl(f(\boldsymbol{x_i};\boldsymbol{\Theta})\bigr)\Bigr]^{\delta_i} \times\\
&\exp\Bigl(-\Lambda_0(T^*_i)\exp\bigl(f(\boldsymbol{x_i};\boldsymbol{\Theta})\bigr)\Bigr)\biggr\}.
\end{split}
\end{equation*}
Moreover, the negative log-likelihood becomes
\begin{equation}\label{fullneg}
\begin{split}
    \mathcal{L}(\Lambda_0,\boldsymbol{\Theta}) = -\frac{1}{n} \sum\limits_{i=1}^n \biggl\{
    &\delta_i \Bigl[ f(\boldsymbol{x_i}; \boldsymbol{\Theta}) +\log\lambda_0(T^*_i) \Bigr] \\
    &- \Lambda_0(T^*_i)\exp\bigl(f(\boldsymbol{x_i};\boldsymbol{\Theta})\bigr)\biggr\},
\end{split}
\end{equation}
%$$\ell(\Lambda_0,\Theta) = -  \sum\limits_{i}\left[\delta_i \left( f(x_i; \Theta) +\log\lambda_0(T^*_i) \right) -  e^{f(x_i; \Theta)}\Lambda_0(T^*_i)\right],$$
which depends on both $\Lambda_0$ and parameters $\boldsymbol{\Theta}$ in $f$.

In practice, the prior knowledge of $\Lambda_0$ is not available. To overcome this issue, we adopted the non-parametric Breslow estimator, which treated the baseline as a piece-wise constant between event failure times:
\begin{equation*}
\begin{split}
\hat{\Lambda}_0(t; \boldsymbol{\Theta}) &= \sum\limits_{j \notin R(t)}\Delta\hat{\Lambda}_0(T^*_j) \\
&=  \sum\limits_{j \notin R(t)}\frac{\delta_j}{\sum\limits_{k\in \mathcal R(T^*_j)} \exp{\bigl( f(\boldsymbol{x_k};\boldsymbol{ \Theta})\bigr)}}.
\end{split}
\end{equation*}

Plugged it into the negative log-likelihood Eq.\ref{fullneg}, we derived the partial likelihood without $\lambda_0(t)$:
\begin{equation}\label{fbloss}\begin{split}
    \mathcal{L}_{\rm fb}(\boldsymbol{\Theta})= -\frac{1}{n} \sum\limits_{i} \delta_i\biggl[ f(\boldsymbol{x_i}; \boldsymbol{\Theta}) - \log\sum\limits_{j\in  \mathcal R(T^*_i)} \exp{\bigl(f(\boldsymbol{x_j}; \boldsymbol{\Theta}) \bigr)}  \biggr]. 
\end{split} 
\end{equation}

%\tlee{don't use things like "We'll" in research papers, always do "We will"} \xwang{fixed}

We refer to this as the \textit{full-batched loss} in this paper. In fact, the procedure of getting partial likelihood of the Cox proportional model can lead us to the equivalent loss function. Given the model $\lambda(t) =\lambda_0(t)\exp\bigl(f(\boldsymbol{x};\boldsymbol{\Theta})\bigr) $,  the partial likelihood now becomes 

\begin{equation}
\label{eq:partial_likelihood}
L(\boldsymbol{\Theta})_{\rm partial} = \prod_{i =1}^{n} \biggl[\frac{\exp\bigl(f(\boldsymbol{x_i};\boldsymbol{\Theta})\bigr)}{\sum\limits_{j\in R(T^*_i)}\exp\bigl(f(\boldsymbol{x_j};\boldsymbol{\Theta})\bigr)}\biggr]^{\delta_i},
\end{equation}
The full-batched loss function can be obtained by taking the average of the negative log of the partial likelihood.  
%\tlee{this sentence is not finished?} \xwang{fixed}

Even though the full-batched loss is convex in $f$, due to the non-convexity of the neural network, the full-batched loss is non-convex. Also, the full-batched loss involves complicated sums over the risk set, which can be as large as the full data set, making it computationally expensive.
    
To deal with the non-convexity and make it scalable to large datasets, we modified the full-batched loss by first subsampling the data and collecting them to a batch $\Omega$, and then restricting the risk set $R(T^*_i)$ only to contain the subsampled data in the current batch: 
\begin{equation}\label{mbloss}
    \tilde{\mathcal{L}}_{\rm mb}(\Theta)= -\frac{1}{|\Omega|} \sum\limits_{i\in \Omega} \delta_i\biggl[ f(\boldsymbol{x_i}; \boldsymbol{\Theta})- \log\sum_{j} \exp{\bigl(f(\boldsymbol{x_j}; \boldsymbol{\Theta}) \bigr)}  \biggr]
\end{equation}
with $j\in\mathcal R(T^*_i)\cap\Omega$. We refer to this expression as the \textit{mini-batched loss} in the paper. If we set the batch as the full data set, then the mini-batched loss is equivalent to the full-batched loss. The batch size can be as small as 2. By restricting data to a randomly sampled batch, we avoided massive calculations. The mini-batched loss is unlike the minibatch gradient descent with i.i.d. data with respect to the full-batched loss since taking the expectation over random minibatch samples does not give the averaged negative log-likelihood. 

As an aside, we can see that the partial likelihood in \eqref{eq:partial_likelihood} is the likelihood of observing the given order of events, which in this case is the order of individuals' deaths.
By evaluating the partial likelihood, we are in effect ignoring any information of the timing of the events beyond just their ordering.
This objective and the mini-batch gradient descent described above appear in recommendation system applications where user preferences are expressed via the relative ordering of click-through events.
The resulting method is called listwise ranking in the recommendation system literature \cite{cao2007learning, wu2018sql}.

%\xwang{XW: Is the last sentence necessary?}
%\tlee{$mb$ under $L$ should be ${\rm mb}$.  I think we can keep the last sentence.} \xwang{fixed}

\subsection{Two-task Method for Disease Diagnosis and Survival Hazard Prediction} 

Lung cancer is one of the most common cancers. Computed Tomography (CT) images, which include a series of axial image slices that visualize the tissues and nodules within the lung area, can be extremely useful for diagnosis purposes. When given a patient's pulmonary CT images, one objective is to diagnose whether the patient has lung cancer or not, i.e., lung cancer classification. In addition, we hope to predict the severity of cancer by estimating the patient's risk of dying from lung cancer, i.e., survival hazard prediction. Traditionally, to fulfill the two tasks, one option is to train separate models with different losses, respectively: binary cross entropy for lung cancer classification and mini-batched loss for survival hazard prediction. However, it raises concerns about divergent predictions, which may result in predicting a case without lung cancer but with a high risk of mortality of dying from lung cancer. 

The link between lung cancer diagnosis and survival prediction is established through the comprehensive analysis of imaging studies. Extracted information from CT images, such as the presence of lung nodules and detailed characteristics (including size, shape, location, and tumor spread), is not only instrumental in confirming the presence of cancer, but also provides critical details that inform prognosis, guide treatment decisions, and influence survival predictions for individual patients. The higher the probability of having lung cancer inferred from CT images, the more likely it is that the cancer exhibits features associated with an advanced or aggressive nature. These features contribute to an increased risk of mortality, forming the basis for the correlation between the probability of having lung cancer and survival prediction. The integration of imaging data into a holistic approach enhances the precision and personalized nature of lung cancer care. 

\begin{figure}[htbp]
    \centering
    \includegraphics[width=3.5in]{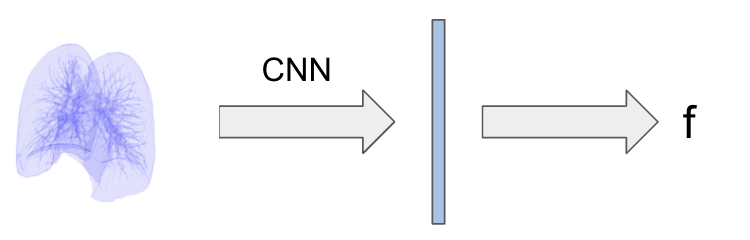}
    \caption{{\bf Two-task Convolution Neural Network Illustration.}}
    \label{2op}
\end{figure}

Recognizing the clinical need to integrate these tasks, we present a novel method capable of simultaneously performing lung cancer classification and survival hazard prediction using the same input – a two-task neural net framework, as illustrated in Fig ~\ref{2op}. The output layer, which predicted the log relative hazards $f(\boldsymbol{x}; \boldsymbol{\Theta})$, was also used for lung cancer classification with sigmoid activation. This choice is intuitive as the function $f$ represents hazard, implying that a higher hazard is indicative of a higher probability of having lung cancer. Instead of having separate losses, we defined the loss as the sum of binary cross entropy and the batched loss. Let $y_i$ be the indicator of having lung cancer, $\boldsymbol{x_i}$ be the image input to the deep neural network, and  \ $f(\boldsymbol{x_i};\boldsymbol{\Theta}) $ be the neural network output for log relative hazards, $P(\boldsymbol{x_i};\boldsymbol{\Theta}) = sigmoid(f\bigl(\boldsymbol{x_i};\boldsymbol{\Theta})\bigr)$ is predicted cancer probability: 
\begin{equation}
\begin{split}
         L(\Theta) = & -\frac{1}{|\Omega|} \sum\limits_{i\in \Omega}\biggl\{\delta_i\Bigl[ f(\boldsymbol{x_i}; \boldsymbol{\Theta}) - \log\sum\limits_{j} \exp{\bigl(f(\boldsymbol{x_j}; \boldsymbol{\Theta}) \bigr)}  \Bigr] \\
         & +\Bigl[ y_i\log P(\boldsymbol{x_i};\boldsymbol{\Theta})+(1-y_i)\log \bigl( 1- P(\boldsymbol{x_i};\boldsymbol{\Theta}) \bigr) \Bigr]\biggr\},
\end{split}
\end{equation}
with $ j\in  \mathcal R(T^*_i)\cap\Omega$.

% Place figure captions after the first paragraph in which they are cited.

One advantage of this approach is consolidating the goals of cancer classification and survival hazard prediction into a singular model, motivated by the clinical reality that the CT image shows information that is critical for both cancer diagnosis and survival prediction. Training a unified model concurrently for both objectives with shared neural net parameters promises a more comprehensive understanding and superior predictive performance, while conventional approaches of training separate models with binary cross entropy for cancer classification and mini-batched loss for hazard prediction focus exclusively on one aspect. This two-task method provides a holistic view, bridging the diagnostic and prognostic aspects of lung cancer, and offers a more clinically relevant perspective for personalized patient care decisions. Another advantage lies in the dual losses, which enable more comprehensive supervision of the neural net's fit, thereby preventing overfitting during training.

\subsection{Evaluation Metrics}

For the cancer classification task, we used AUC (area under the ROC curve) to evaluate the model performance. In the hazard prediction task, we employed the concordance index (C-index) for evaluation. C-index, introduced by Harrell \textit{et al.} \cite{harrell1982evaluating}, is a goodness of fit measure for models that produce risk scores for censored data.  In our context, it estimates the probability that, for any random pair of individuals, the predicted survival times would exhibit the same ordering as their actual survival times. This is equivalent to determining whether, for any random pair of patients, the predicted hazard has the reverse order in comparison to their actual survival times, as patients with higher predicted survival times correspond to lower predicted hazards.
%\tlee{I don't quite understand the goal of this sentence: "Boiling this intuition down to two comparable patients: the patient with the higher hazard score should have a shorter time-to-event."} \xwang{ the C-index expression below uses the inverse relationship between risk and survival time to approximate the concordance, the $\approx$ part. Rephrased. } 
The C-index in our context is defined by the following formula:
\begin{eqnarray}
C &=& \frac{\text{\# concordant pairs}}{\text{\# concordant pairs + \# disconcordant pairs}} \nonumber \\
       &=& P\{\hat{T_i}> \hat{T_j}|\ T_i>T_j,\delta_j =1\} \nonumber \\
\label{eqn:approx}
       &\approx& P\{\hat{f_i}< \hat{f_j}|\ T_i>T_j,\delta_j =1\} \\
&=& \frac{\sum_{i\neq j}1\{\hat{f_i}< \hat{f_j}\}1\{T_i>T_j\}\delta_j}{\sum_{i\neq j}1\{T_i>T_j\}\delta_j}, \nonumber
\end{eqnarray}
where approximation~(\ref{eqn:approx}) follows from the argument that a patient with a higher hazard score should have a shorter survival time.

When $\mbox{C-index}=1$, it corresponds to the scenario where the order of the predictions is the same as that of the true survival times, while $\mbox{C-index}=0.5$ represents a random prediction. Typically, a model with a C-index above 0.7 can be regarded as a good model.

\section{Simulation Studies}\label{sec:simu}
% Three simulations were conducted to investigate our proposed method empirically. 
 This section reports results from three simulation experiments.
 Both Simulations A and B focused on the extended Cox model and its prediction of the log relative hazards function $f$. Simulation A was under the setting where there were event cases only, while Simulation B involved both censored and event cases. Both simulations used the same images from the MNIST dataset and the same generated survival time, but different censoring statuses. We compared the performance of the oracle loss, full-batched loss, and mini-batched loss under the settings of Simulations A and B. Simulation C was designed for the two-task framework, involving both the disease occurrence classification and the survival hazard prediction with the log relative hazards function. We generated a new dataset from the CIFAR-10 dataset, called Nodule-CIFAR. We compared the loss function performance of the combination of binary cross-entropy and full-batched/mini-batched in terms of AUC and C-index.

\subsection{Simulations A and B}
\subsubsection{MNIST Dataset and Time-to-event Data}
\begin{figure}[htbp]
    \centering
    \subfloat[Survival Time Distribution without Censoring]{
        \includegraphics[width=3.5in,height=2in]{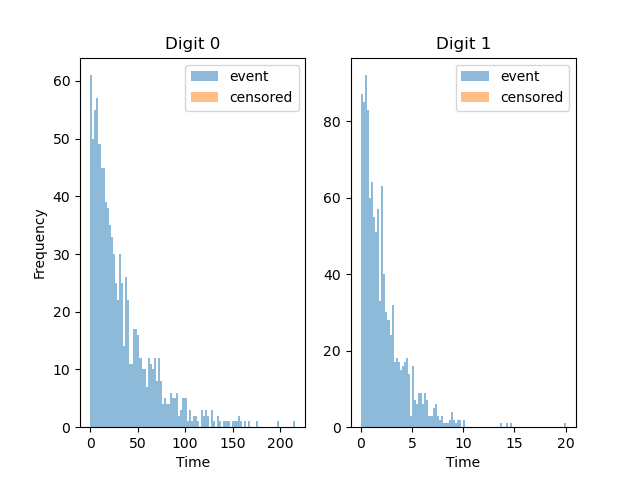}
        \label{subfig:a}
    }
    \vfill
    \subfloat[Survival Time Distribution with Censoring]{
        \includegraphics[width=3.5in,height=2in]{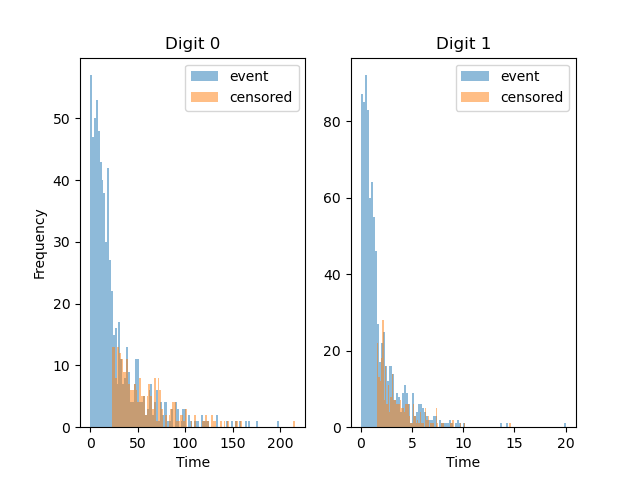}
        \label{subfig:b}
 }
    \caption{{\bf Simulated Survival Time Distributions}. (a) Survival time distributions for the two digits in Simulation A, without the censoring mechanism; (b) Survival time distributions for the two digits in Simulation B, with the censoring mechanism. The censored cases are labeled in orange, which overlaps the upper half of the event cases.}
    \label{hist}
\end{figure}

We used the MNIST image dataset and generated artificial survival times for digits in our simulations. The MNIST dataset is an image dataset of handwritten digits from 0 to 9; see \cite{deng2012mnist}. We selected 2 digits from the MNIST dataset as input images of the neural network with different patterns, w.l.o.g., we selected zeros and ones. We generated the survival time for each digit using an exponential distribution with different constant hazards $\lambda_j = 1 \times \exp(\phi_j)$, $j = 0, 1$, where the baseline hazard $\lambda_0(t) $ was set to 1, and the true log relative hazards was $\phi_j$. In Simulation A, all cases were labeled as events. In Simulation B, we randomly labeled half of the individuals who lived beyond the median as censored cases within each digit. The distribution of the test set is shown in Fig~\ref{hist}.

\subsubsection{Architecture}
Simulations A and B were trained under the same feed-forward convolution neural network, which consisted of a stack of convolution and dense layers. The net structure is listed in Table~\ref{table1}.
\begin{table}[!t]
\caption{Convolution Neural Net Architecture for Simulations A \& B \label{table1}}   
    \centering
    \scalebox{0.9}{
    \begin{tabular}{  c c c c }
    \hline
     Layer Type& Number of Kernels  & Kernel Size & Output Size\\ 
    \hline
    Convolution &32 & $5\times5$ & $28\times28\times32$ \\
    Max Pooling &&$2\times2$, stride = 2& $14\times14\times32$ \\
    Convolution &64 & $5\times5$ &  $14\times14\times64$\\
    Max Pooling &&$2\times2$, stride = 2& $7\times7\times64$ \\
    Flatten &&& 3136\\
    Fully Connected &&& 1024\\
    Fully Connected &&& 128\\
    Fully Connected &&& 1\\
    \hline
    \end{tabular}
    }
\end{table}
    
\subsubsection{Results of Simulations A and B}
We introduced the oracle loss (see Eq \ref{orcloss} and \ref{orcbatchedloss}). It leverages the prior knowledge of the baseline hazard $\lambda_0(t)$ when compared with the full-batched loss~(Eq \ref{fbloss}) and mini-batched loss~( Eq \ref{mbloss}), i. In our simulations, w.l.o.g., we set $\lambda_0(t) = 1$ when generating survival time, so that $\Lambda_0(t) = t$. Plugging the baseline hazard into the averaged negative full log-likelihood~(Eq \ref{fullneg}) provided us the oracle loss, for which $f$ can be trained:
\begin{equation}\label{orcloss}
    \mathcal{L}_{\rm orc}(\Theta) = -\frac{1}{n}
        \sum\limits_{i=1}^n
        \Bigl[\delta_i f(\boldsymbol{x_i};\boldsymbol{\Theta}) - \exp\bigl(f(\boldsymbol{x_i};\boldsymbol{\Theta})\bigr)T^*_i
        \Bigr].
\end{equation}

Due to the non-convexity of neural network $f$, we used the stochastic gradient descent (SGD) method to minimize the non-convex loss function. Correspondingly, the batched version is provided below.
\begin{equation}\label{orcbatchedloss}
\tilde{\mathcal{L}}_{\rm orc}(\Theta) =- \frac{1}{|\Omega|} 
        \sum\limits_{i\in\Omega} \Bigl[\delta_i f(\boldsymbol{x_i};\boldsymbol{\Theta}) - \exp\bigl(f(\boldsymbol{x_i};\boldsymbol{\Theta})\bigr)T^*_i
        \Bigr],
\end{equation}
where $\Omega$ is the selected batch for a training iteration. We will later refer to this as the \textit{oracle loss}. 

%The reason we can set base rate $\lambda_0(t)$ to 1: The inverse function $g$ of cumulative base rate hazard $\Lambda_0(t)$ exists because of its monotonic property. With the transformation of time $t = g(\tau)$,  $\tilde{\Lambda}_0(\tau) =\Lambda_0(g(\tau)) = \tau $, and $\tilde{\lambda}_0(\tau) = 1$. 
%\tlee{do we need this paragraph?}\xwang{removed}

We also calculated the true loss as the baseline for benchmark comparisons. When both the baseline hazard $\lambda_0(t)$ and the log relative hazards $\phi_j$ were available, we could directly plug them into the averaged negative full log-likelihood~(Eq \ref{fullneg}), which gave the true loss.

 \begin{table}[!t]
     \caption{Simulations A \& B: C-indexes under three losses\label{simuAB-Cind}}
    \centering
    \begin{tabular}{  c| c c c }
    \hline
      & Oracle & Full-batched  & Mini-batched \\
      \hline
      A & 0.7268 &0.7165 &0.7189\\
      \hline
      B w/ censored (C1) & 0.7184 & 0.7146 & 0.7166\\
      \hline
      B w/o censored (C2) &0.6845&0.6770& 0.6790\\
    \hline
    \end{tabular}
\end{table}
    
\begin{figure}[!h]
    \centering
    \subfloat[Simulation A]{
        \includegraphics[width=2.5in,height =2in]{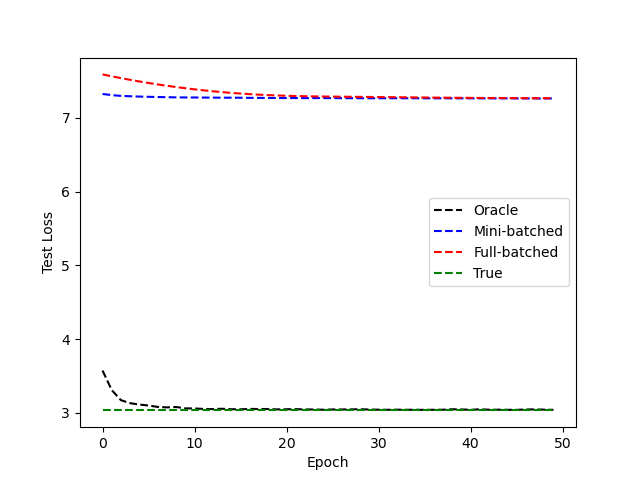}
    }
    \hfill
    \subfloat[Simulation B]{
       \includegraphics[width=2.5in,height =2in]{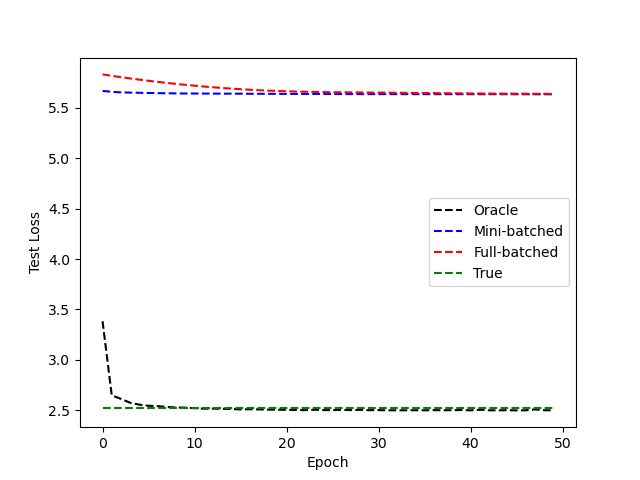} 
     } \\  
    \caption{ \bf {Simulation Losses by Epoch}.  (a) Simulation A (b) Simulation B.}
    \label{simuAB-loss}
\end{figure}

Results of Simulations A and B are reported in Fig~\ref{simuAB-loss} and Table~\ref{simuAB-Cind}. In both simulations, the oracle loss settled to the true loss, the oracle loss was less than the batched losses, both batched losses settled to the same value, and the mini-batched loss settled faster than the full-batched loss. This met our expectations since the oracle loss had access to the base rate. In addition, due to the extra information, the C-index trained by the oracle loss is expected to be larger, which was validated in both Simulations A and B, see Table~\ref{simuAB-Cind}. In Simulation A, though the C-index curve fluctuated after loss converges, it achieved a high value for both full batched loss and mini-batched loss, showing good rank prediction on the hazards when there is no censoring. In Simulation B, two C-indexes were calculated: $C_1$ involved both censored and event case, while $C_2$ involved event cases only. Here, $C_1$ exceeds 0.7, which means good rank predictions for pairs across censored and event groups and pairs within the event group. Moreover, the faster convergence and small difference between $C_{orc}$ and $C_{mb}$ indicated the feasibility of mini-batched loss for training parameters without prior information of $\lambda_0(t)$.
    
\subsection{Simulation C: Nodule-CIFAR Simulation with Classification and Hazard Prediction}

\subsubsection{Nodule-CIFAR Dataset} We introduced a new dataset, called Nodule-CIFAR, which was generated from the CIFAR-10 dataset \cite{krizhevsky2009learning}. Nodule-CIFAR was inspired by Tumor-CIFAR from Gao \textit{et al.} \cite{gao2019distanced} and simulated benign and malignant nodules on the CIFAR-10 images. In reality, benign nodules typically exhibit smaller sizes with regular round shapes and are non-cancerous, while malignant nodules tend to be larger in size and exhibit irregular shapes. Healthy individuals possess benign nodules, but patients may have both benign and malignant nodules. To simulate this, we introduced black and white dots onto CIFAR-10 images to simulate benign nodules, while dummy nodules were represented as white blobs to simulate malignant nodules. 

\begin{figure}[htbp]
\centering  
\includegraphics[width=3.5in,height =2in]{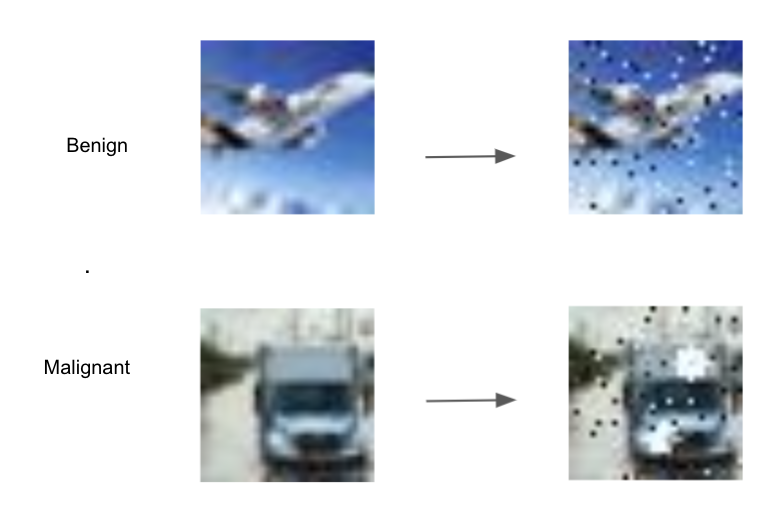}
\caption{\bf {A Nodule-CIFAR Example}. Non-cancer cases only have small black and white dots scattered over the images, simulating benign nodules. In addition to benign nodules, cancer cases have 2 larger white patches to simulate malignant nodules.}
\label{nodule-CIFAR}
\end{figure}

There were 10,000 samples in the training set and 1,000 samples in the testing set. We randomly assigned images to non-cancerous and cancerous groups with equal probability, so that cancer prevalence was 50\% in both training and test sets. Among the cancerous cases, we randomly labeled 50\% as censored, and the remaining were labeled as events, the events of failure of dying from cancer. For the non-cancerous cases, they would not die of cancer, so all of them were labeled as censored. Next, we incorporated simulated nodules, either benign or malignant, onto CIFAR-10 images based on their assigned group. The non-cancer images yet featuring benign nodules, displayed numerous small black and white dots distributed across the image to simulate benign nodules. In contrast, the images in the cancer groups had two additional big white patches randomly located in the images, mimicking malignant nodules. Within the cancer group, the censored had relatively smaller white patches compared to the event, because the censored group had not yet reached a deadly stage. The original image categories from the CIFAR-10 dataset were irrelevant in this context; the distinctions between cancer and non-cancer were determined by the presence of simulated white patches. Moreover, within the cancer group, the censoring status was solely associated with the sizes of the simulated white patches. Fig~\ref{nodule-CIFAR} is an example of images in the Nodule-CIFAR dataset.

Time-to-event data corresponding to Nodule-CIFAR images were generated based on the largest size of simulated nodules in each image. The recorded time followed an exponential distribution with a parameter of $\lambda = 1 \times \exp(\phi)$, where $\phi \propto size$, the largest size of simulated nodules in each image. This was consistent with our expectation that the larger the nodule size, the larger the hazards, and the smaller the survival time. 

\begin{figure}[htbp]
    \centering
    \subfloat[ Nodule Size Distribution]{
        \includegraphics[width=2.5in,height =2in]{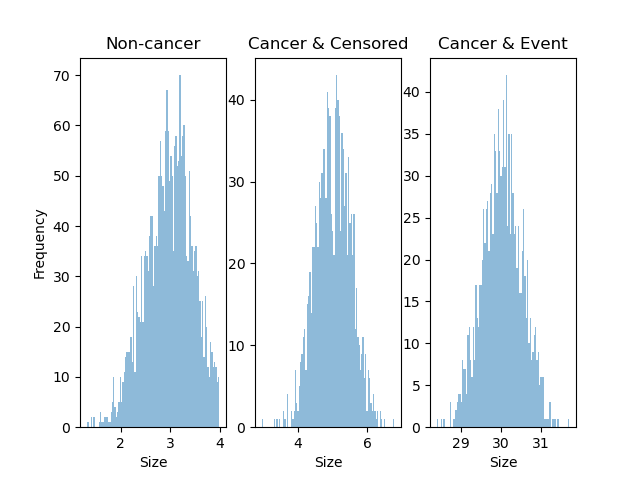}
    }
       \vspace*{-.15in}
    \hfill
    \subfloat[    Survival Time Distribution]{
        \includegraphics[width=2.5in,height =2in]{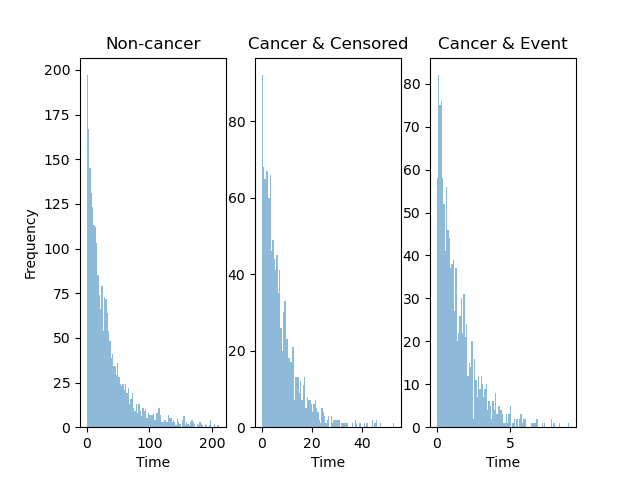} 
     }\\ 
     \vspace{0.2in}
   \caption{\bf {Nodule Size and Survival Time Distribution by Group}. (a) Nodule size distribution by group. The non-cancer group has smaller nodules on average when compared with the cancer group. Within the cancer group, event cases (those who eventually die of cancer in simulation) have larger malignant nodules. (b) Survival time distribution by group in Nodule-CIFAR. The time-to-event for the non-cancer group is larger than the cancer group. Within the cancer group, the time-to-event of censored is larger than that of the event cases.}
   \label{hist-nodule-CIFAR}
\end{figure}

Fig~\ref{hist-nodule-CIFAR} shows the distribution of nodule size and survival time for each group. The non-cancer group had smaller nodules on average compared to the cancer group. Within the cancer group, those event cases (eventually died of cancer in simulation) had larger malignant nodules. The time-to-event for the non-cancer group was larger than the cancer group. Within the cancer group, the time-to-event of censored cases was larger than the event cases. 

\subsubsection{Architecture}

Like Simulations A and B, Simulation C was trained under a feed-forward convolution neural network, which consisted of a stack of convolution and dense layers. The output was used for both disease occurrence classification and hazard prediction evaluation. See Table \ref{nc-net} for the structure of the neural network.

 \begin{table}[!t]
     \caption{Convolution Neural Net Architecture for Simulation C
    \label{nc-net}}
    \centering
    \scalebox{0.7}{
    \begin{tabular}{  c c c c }
    \hline
     Layer Type& Number of Kernels  & Kernel Size & Output Size\\ 
    \hline
    Convolution &32 & $5\times5$ & $28\times28\times32$ \\
    Max Pooling &&$2\times2$, stride = 2& $14\times14\times32$ \\
    Convolution &64 & $5\times5$ &  $14\times14\times64$\\
    Max Pooling &&$2\times2$, stride = 2& $7\times7\times64$ \\
    
    Flatten &&& 3136\\
    Fully Connected &&& 100\\
    Fully Connected &&& 10\\
    Fully Connected &&& 1\\    
    \hline
    \end{tabular}}
    \end{table}

\subsubsection{Results of Simulation C}
\begin{figure}[!h]
    \centering
    %\vspace*{-.02in}
    \subfloat[Test Loss]{
        %\vspace*{-.4in}
        \includegraphics[width = 2.5in]{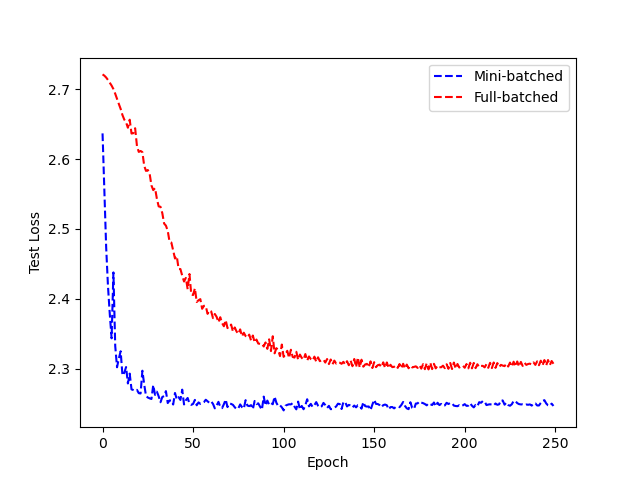}
        \label{simucloss}
    }
       %\vspace*{-.1in}
    \hfill
    \subfloat[AUC]{
        %\vspace*{-.2in}
        \includegraphics[width = 2.5in]{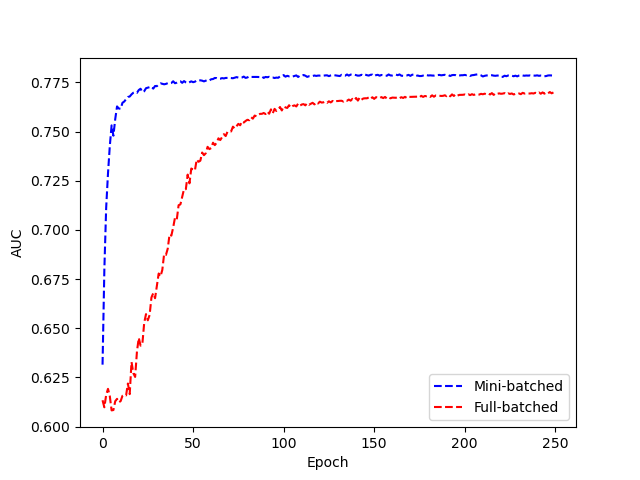}
        \label{simucauc}
    }
       %\vspace*{-.1in}
    \\
    \subfloat[C-index for All Cases]{
        %\vspace*{-.2in}
        \includegraphics[width = 2.5in]{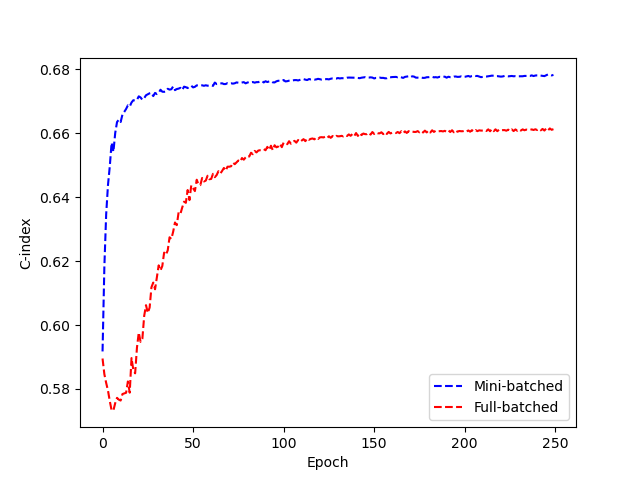 } 
        \label{simucc1}
        %\vspace*{-.1in}
     }
     \hfill
     \subfloat[C-index for Cancer Cases]{
        %\vspace*{-.2in}
        \includegraphics[width = 2.5in]{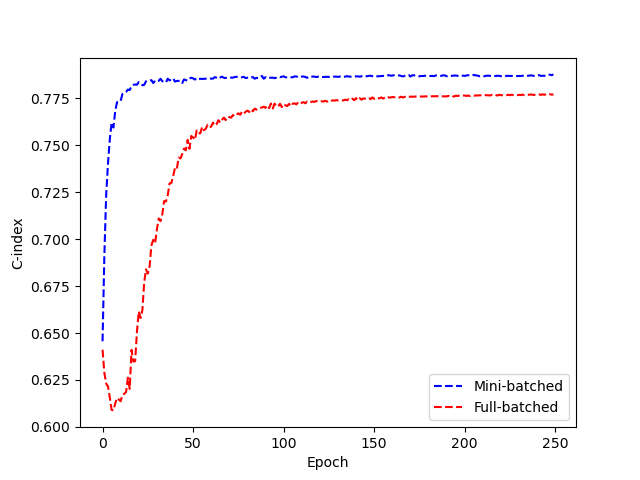 }
        \label{simucc2}
        %\vspace*{-.1in}
     }  \\
     \caption{ \bf{Result of Simulation C}. (a): test loss by epoch; (b): test AUC by epoch; (c): test c-index C1 of all cases by epoch; (d): test c-index C2 of the cancer group by epoch. The sum of binary cross entropy and mini-batched loss performed better in both classification and hazard prediction by achieving higher stabilized AUC, C1, and C2 within fewer epochs. }
    \label{ncresult}
\end{figure}

The loss function for the two-task network was the sum of the binary cross entropy and the full-batched/mini-batched loss. To compare the model performance trained with different losses under the same network architecture, see Fig~\ref{ncresult} for the epoch-wise losses, AUC, and C-index, and Table~\ref{simuC} for their stabilized values after the losses converge. As shown in Fig~\ref{simucloss}, the one with mini-batched loss (blue) converged much faster than the one with full-batched loss (red); it reached a minimum after a few epochs and stabilized. Fig~\ref{simucauc} showed both losses outperformed the baseline AUC 50\% significantly, which was achieved by predicting all cases as non-cancer, and the model trained with mini-batched loss achieved a slightly higher AUC. As for the hazard prediction evaluation, we calculated two C-indexes $C1$ and $C2$, where $C1$ was for all cases (cancer and non-cancer, Fig~\ref{simucc1}) and $C2$ was for the cancer group (Fig~\ref{simucc2}). Both losses achieved competitive $C1$ and $C2$ values, especially within the cancer group, where $C1$ exceeded 0.75 for both losses. Comparing Fig~\ref{simucc1} and Fig~\ref{simucc2}, we noticed the C-index decreased to around 0.65 when it involved the non-cancer group, which was caused by the trade-off between the classification and hazard prediction tasks. Overall, the sum of binary cross entropy and the mini-batched loss performed better in both classification and hazard prediction by achieving higher stabilized AUC and C-index values within fewer epochs.

 \begin{table}[!t]
    \caption{Simulation C: AUC and C-index under two losses
    \label{simuC}}
    \centering
    \scalebox{1.0}{
    \begin{tabular}{  c| c c  }
    \hline
      &  Full-batched & Mini-batched \\
      \hline
      AUC & 0.770 & 0.783 \\
      \hline
      C1&0.661 &0.677\\
      \hline
      C2&0.779 &0.785\\
    \hline
    \end{tabular}}
    \end{table}

\section{Real Data Experiment}\label{sec:exp}

\subsection{NLST Dataset}
The National Lung Screening Trial (NLST) collected medical images and survival information from potential lung cancer patients during 2002-2009, see \cite{national2011national}. It was a randomized controlled trial to determine whether screening for lung cancer with low-dose helical computed tomography (CT) reduced mortality from lung cancer in high-risk individuals relative to screening with chest radiography (X-ray). Participants were randomly assigned to two study arms in equal proportions. One arm received low-dose helical CT, while the other received single-view chest radiography.  

CT images are a set of axial slice images of the human body. They can reveal both normal and abnormal tissues inside the organs. The abnormal tissues of the lungs are called nodules. Nodules usually are spherical but may have other shapes. Each sub-type of nodules has a different cancer probability. Hence, doctors take into consideration all nodules when diagnosing lung diseases with CT images.

%In our experiment, we chose the dataset containing medical images of 15,000 patients assigned to the CT group. Among these 15,000 patients, 991 were diagnosed with cancer during the trial period. In our experiment, we selected 991 patients who developed cancer during the trial period from a pool of 15,000 patients who received the CT treatment.  We further collect the latest CT images from 991 patients confirmed with lung cancer, among which 427 died of lung cancer. For the classification task, wechosee the latest CT images from the same number of patients without lung cancer.

%\tlee{only 991 out of 15k developed cancer}\xwang{rephrased}

In our experiment, we chose 991 patients who developed cancer during the trial period from a pool of 15,000 patients who received CT treatment. Subsequently, we collected the most recent CT images from these 991 patients confirmed to have lung cancer, among whom 427 passed away due to lung cancer. For the classification task, we similarly gathered the most recent CT images from an equal number of potential patients who did not have lung cancer. Among the total of 1882 patients, those with confirmed lung cancer cases were assigned a label of $y_i = 1$, while all others were labeled as $y_i = 0$. In addition, those who experienced lung cancer-related mortality were categorized as events of failure (non-censored) with $\delta_i = 1$, whereas the rest were considered censored with $\delta_i = 0$. Each patient's most recent CT examination was utilized as the input image denoted as $X$. Furthermore, we collected patients' survival time $T^*$ by subtracting their latest exam date from the date they were last known alive.

%We preprocess all of the selected images by the method introduced in the method section. The input of the network is the top 5 suspicious nodule crops. 

\subsection{Preprocessing}
In terms of preprocessing the CT images from NLST datasets, we utilized the open-source code \cite{zuidhof_2017} to segment the lungs from the CT images and applied the nodule detection method described in \cite{liao2019evaluate} to obtain the top 5 suspicious nodule crops as input. For completeness, we provide a brief summary of their method below.

\subsubsection{Lung Segmentation}
The CT images are a set of cross-sectional images of the body. Preprocessing for lung segmentation was required before they were ready for the CNN. First, the CT scans should be resampled to $1 \times 1 \times 1 mm^3$ isotropic resolution, then the resampled CT scans were preprocessed with the following main steps:
\begin{enumerate}
\item [i.] Mask extraction: The first step was to extract the lungs' mask by converting the image to Hounsfield unit (HU) and binarizing the image with the lungs' HU values. HU is a standard quantitative scale for describing radiodensity. Each organ has a specific HU range, and the range remains the same for different people. Here, we used a $-320$ HU value as the threshold for the lungs. The largest connected component located in the center of the image was extracted as the lungs' mask.
%\tlee{what is the threshold?} \xwang{added}

\item [ii.] Convex hull computation: The second step was to compute the convex hull of the lungs' mask. Because some nodules might be connected to the outer lung wall and might not be covered by the mask obtained in the previous step, a preferred approach was to obtain the convex hull of the mask. However, it could include other unrelated tissues if one directly computes the convex hull of the mask. To overcome this issue, we first divided the mask into left and right lung masks, then computed their respective convex hulls, and lastly merged them to form the final, whole lungs' convex hull.  
%\tlee{please check if the new writing is okay}

%\tlee{not clear; also, need a step name?}\xwang{rewrote and removed the first step name}

\item [iii.] Lung segmentation: We obtained a segmentation of the lungs by first multiplying the CT image with the mask and then filling the masked region with tissue luminance. 
\end{enumerate}

\begin{figure}[htbp]
\centering
\includegraphics[height = 3in,width = 3in]{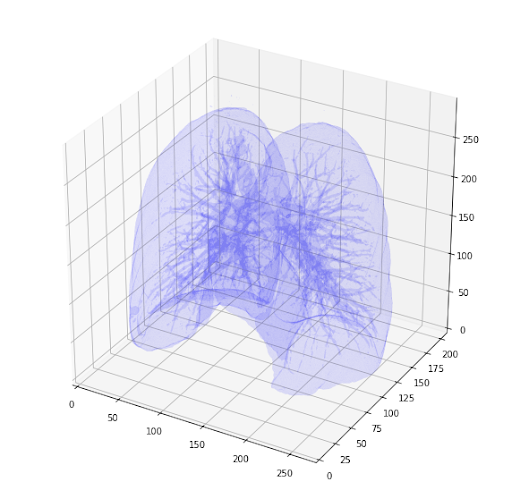}
\caption{\bf {An Example for 3D Segmented Lungs from CT images.}}
\label{preprocess}
\end{figure}

After completing these three steps, 3D segmented lungs can be extracted.  An example is shown in Fig~\ref{preprocess}. 
%\tlee{why the reference?} \xwang{removed the preprocess step plot}

\subsubsection{Nodule Detection}

The sizes of the segmented lung images varied for each patient, which went against the requirement for identical image sizes in CNNs to work properly. To resolve this, the segmented images were resampled to the same resolution and fixed slice distance. Although the size of each cropped image might differ due to varying lung sizes among patients, zero padding was used if the image size is less than $224\times224\times224\times1$; otherwise, the central 224-width cubes were extracted. 
An attempt was made to directly input this preprocessed 224-cube into a 3D network for lung cancer classification and hazard prediction. Still, it was computationally time-consuming, and the results were unsatisfactory due to the large size of 3D images and potential memory issues. To address the issue, we followed Liao \textit{et al.} 's nodule detection process \cite{liao2019evaluate}. The nodule detector took in the 3D segmented lung CT image and output predicted nodule proposals with their center coordinates, radius, and confidence. The five most suspicious lung proposals were selected as input $X$ for our network, as Liao \textit{et al.} determined that $k=5$ was sufficient for recall when different top $k$ proposals with the highest confidence were selected for inference \cite{liao2019evaluate}. For each selected proposal, a $96 \times 96 \times 96 \times 1 $ patch centered on the proposed nodule was cropped, resulting in an input size of $5 \times96 \times 96 \times 96 \times 1 $, where one channel represented the number of channels.

\subsection{Network Structure}
\begin{figure}[!t]
\centering  
\includegraphics[height = 2.5in,width = 3in]{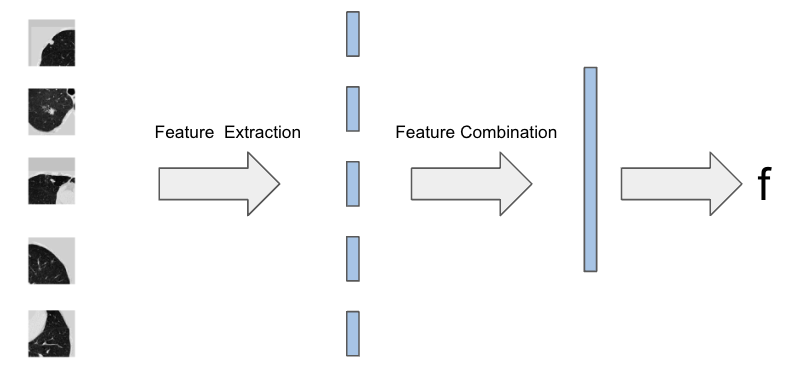}
\caption{\bf {Network structure with 2 phases}. Convolution and integration phases.}
\label{2outputs}
\end{figure}

The top five regions with the highest nodule confidence were considered for cancer occurrence classification and hazard prediction tasks for each patient. The network had two phases: feature extraction from each lung crop using convolutional layers, and feature combination through the integration, as shown in Fig~\ref{2outputs}. The final output $f$ was evaluated with AUC and C-index metrics.

\subsubsection{Convolution Phase}

We had three different convolution structures to extract features from the top five nodule crops: Alex3D, VGG163D, and Res-net18. Each took a nodule proposal as input and output a 128-D feature. We also adopted the pre-trained cancer classifier from Liao \textit{et al.} \cite{liao2019evaluate} as a performance benchmark.

\subsubsection{3D Alex Net}
Table \ref{3dalex} lists layers in Alex 3D. The network was based on the classic 2D Alex Net architecture with modifications specifically tailored for the NLST dataset.
 \begin{table}[h!]
 \caption{3D Alex Net architecture for lung CT images\label{3dalex}}
\centering
\scalebox{0.9}{
\begin{tabular}{  c c c c }
\hline
 Layer Type& Number of Kernels  & Kernel Size & Output Size\\ 
\hline
Convolution &96 & $3\times3\times3$ & $48\times48\times48\times96$ \\
Max Pooling && $3\times3\times3$ & $23\times23\times23\times96$ \\
Convolution &256 & $5\times5\times5$ &  $23\times23\times23\times256$ \\
Max Pooling && $3\times3\times3$ &    $11\times11\times11\times256$\\
Convolution & 384 & $3\times3\times3$ &  $ 9\times9\times9\times384$ \\
Convolution &256 & $3\times3\times3$ &  $ 9\times9\times9\times256$ \\
Max Pooling && $3\times3\times3$&   $4\times4\times4\times256$\\
Flatten &&& 16384\\
Fully Connected &&& 4096\\
Fully Connected &&& 128\\
\hline
\end{tabular}}
\end{table}

\subsubsection{3D VGG16}
Table \ref{3dvgg} lists the layers in 3D VGG16 developed from 2D VGG16 \cite{simonyan2014very}, with modifications specifically tailored for the NLST dataset.

\begin{table}[h!]
\caption{3D VGG Net architecture for lung CT images\label{3dvgg}}
\centering
\scalebox{0.9}{
\begin{tabular}{  c c c c }
\hline
 Layer Type& Number of Kernels  & Kernel Size  & Output Size\\ 
\hline
Convolution &64 & $3\times3\times3$ & $96\times96\times96\times64$ \\
Convolution &64 & $3\times3\times3$ & $96\times96\times96\times64$ \\
Max Pooling && $3\times3\times3$ & $48\times48\times48\times64$ \\

Convolution &128 & $3\times3\times3$ & $48\times48\times48\times128$ \\
Convolution &128 & $3\times3\times3$ & $48\times48\times48\times128$ \\
Max Pooling && $3\times3\times3$ & $24\times24\times24\times128$ \\

Convolution &256 & $3\times3\times3$ & $24\times24\times24\times256$ \\
Convolution &256 & $3\times3\times3$ & $24\times24\times24\times256$ \\
Convolution &256 & $3\times3\times3$ & $24\times24\times24\times256$ \\
Max Pooling && $3\times3\times3$ & $12\times12\times12\times256$ \\

Convolution &512 & $3\times3\times3$ & $12\times12\times12\times512$ \\
Convolution &512 & $3\times3\times3$ & $12\times12\times12\times512$ \\
Convolution &512 & $3\times3\times3$ & $12\times12\times12\times512$ \\
Max Pooling && $3\times3\times3$ & $6\times6\times6\times512$ \\

Convolution &512 & $3\times3\times3$ & $6\times6\times6\times512$ \\
Convolution &512 & $3\times3\times3$ & 
$6\times6\times6\times512$\\
Convolution &512 & $3\times3\times3$ & $6\times6\times6\times512$ \\
Max Pooling && $3\times3\times3$ & $3\times3\times3\times512$ \\

Flatten &&& 13824\\
Fully Connected &&& 4096\\
Fully Connected &&& 4096\\
Fully Connected &&& 128\\
\hline
\end{tabular}}

\end{table}

\subsubsection{3D ResNet-18}
Table \ref{3dres} lists the layers in 3D ResNet-18 developed from a 2D residual network \cite{he2016deep}. Downsampling was performed by Res-block2\_1, Res-block3\_1, and Res-block4\_1 with a stride of 2. 

\begin{table}[!t]
\caption{3D ResNet-18 architecture for lung CT images\label{3dres}}
\centering
\scalebox{0.9}{
\begin{tabular}{  c  c c }
\hline
 Layer Name& 3D Resnet-18 & Output Size\\ 
\hline
Conv1 & $7\times7\times7$,64,\text{stride 2} &  $48\times48\times 48\times64$  \\
Max pooling & $3\times3\times3$, \text{stride 2} & $24\times24\times24\times64$ \\
Res-block1 & $\begin{bmatrix}
3\times3\times3, 64 \\
3\times3\times3, 64
\end{bmatrix}$$\times2$ &$24\times24\times24\times64$\\
Res-block2& $\begin{bmatrix}
3\times3\times3, 128\\
3\times3\times3, 128
\end{bmatrix}$$\times2$ & $12\times12\times12\times128$ \\
Res-block3& $\begin{bmatrix}
3\times3\times3, 256\\
3\times3\times3, 256
\end{bmatrix}$$\times2$ & $6\times6\times6\times256$ \\
Res-block4&$\begin{bmatrix}
3\times3\times3, 512\\
3\times3\times3, 512
\end{bmatrix}$$\times2$ & $3\times3\times3\times512$\\
Average-pool& &512 \\
Fully Connected & &128\\
\hline
\end{tabular}
}
\end{table}

\subsubsection{Pretrained Cancer Classifier}
\begin{figure}[htbp]
\centering  
\includegraphics[width = 3in]{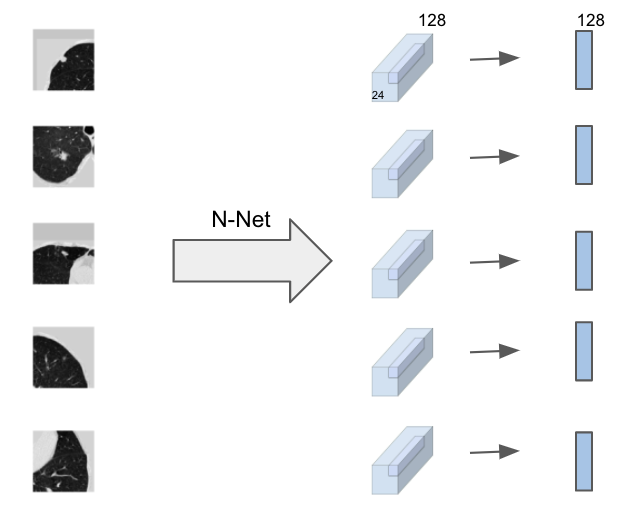}
\caption{\bf { Feature Extraction}. Using pre-trained classifier to get features from top five suspicious crops.}
\label{nnet}
\end{figure}

We adopted the pre-trained cancer classifier from Liao \textit{et al.} \cite{liao2019evaluate} as a performance benchmark. Liao \textit{et al.} \cite{liao2019evaluate} proposes a 3D deep neural network based on U-net for cancer probability reference, which has 2 modules: a nodule detection module and a cancer classification module. Because of the limited data size, the classification module (called N-net) integrates the pre-trained detection module as part of the classifier. We followed Liao \textit{et al.} 's process to obtain the features from image patches: For each selected crop, we fed it to the N-net and obtained the last convolutional layer of the nodule classifier, whose size is $24 \times 24 \times 24 \times 128$. The central $2 \times 2 \times 2$ voxels of each proposal feature were extracted and max-pooled, resulting in a 128-D feature, as shown in Fig~\ref{nnet}.

%\tlee{$\times$ instead of x} \xwang{fixed}

\subsubsection{Integration Phase} 
\begin{figure}[htbp]
\centering  
\includegraphics[width = 3in]{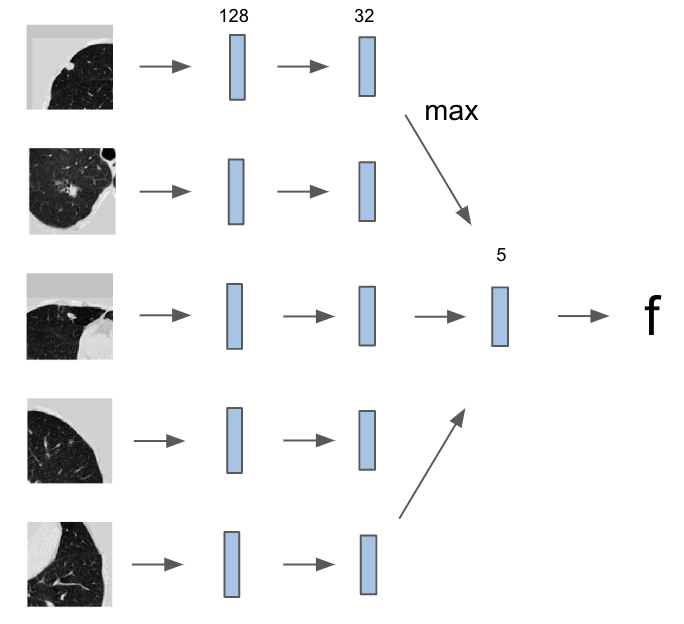}
\caption{\bf {Feature Intergration}. Graphical representation of feature integration process.}
\label{maxp1}
\end{figure}

After the convolution phase, the network had five 128D features for each patient. To obtain a single output from these multiple nodule features, three integration methods were explored. The best-performing integration method is shown in Table~\ref{table5}, and its graphical representation can be found in Fig~\ref{maxp1}. The features from the top five nodules were individually input into a fully connected layer with 32 hidden units. The maximum value of each feature was considered for the final result after concatenating into a single 5D feature, and a following fully connected layer generated the final output $f$. 

\begin{table}[!h]
\caption{Integration Phase Structure\label{table5}}
    \centering
    \scalebox{1}{
    \begin{tabular}{  c  c }
    \hline
     Layer Type& Output Size\\ 
    \hline
    Convolutional Phase Output & $128 \times 5$\\
    Fully Connected & $32 \times 5$\\
    Max Pool & $1 \times5$ \\
    Fully Connected & 1\\
    \hline
    \end{tabular}
    }
\end{table}

%the cancer probability of each nodule. The maximum of these probabilities is then taken as the probability $P$ of the patient since the presence of at least one malignant nodule would indicate that the patient has cancer. The AUC and c-index, however, use the fully connected layer outputs $f$, as shown in Fig.\ref{maxp1}.

\subsection{Results}
The AUC of the lung cancer occurrence classification and the C-index of the hazard prediction are listed in Table~\ref{result} for the pair-wise combination of four convolution methods and one integration method. The C-index was calculated based on both cancer and non-cancer groups, as the non-cancer group in the NLST data set were individuals who had the potential risk of developing cancer. Compared to the pre-trained network of \cite{liao2019evaluate}, all three architectures achieved higher AUC and C-index values, indicating better lung cancer classification and survival prediction. 

\begin{table}[!h]
\caption{Results of NLST Experiment\label{result}}
\centering
\scalebox{1}{
\begin{tabular}{   c| cc }
\hline
   &AUC &C-index \\ 
\hline
3D Alex   &  0.674 &0.601 \\
\hline
3D Res18   & 0.690 & 0.601 \\
\hline
3D VGG16    & 0.680 & 0.608 \\
\hline
Pretrained & 0.550 & 0.519\\
\hline
\end{tabular}}
\end{table}

\section{Discussion}

The results of our study suggest that the combination of the binary cross-entropy and mini batched loss, obtained by extending the Cox model with CNN, holds the potential to improve the diagnosis and treatment of lung cancer. Our approach demonstrates a high AUC in lung cancer classification and a high C-index in survival prediction, using CT images from the NLST dataset. One strength of our approach is the use of the mini-batched loss, which effectively handles the non-convexity induced by neural networks and enables the training of large datasets. Additionally, the combination of the mini-batched loss with binary cross-entropy allows for both lung cancer classification and survival hazard prediction. Furthermore, this approach has the potential to be generalized with any type of medical images beyond CT scans. A model can be trained with medical images along with corresponding survival time information to predict the disease occurrence and risk of mortality.   

%Our study has several limitations that need to be addressed in future research. Firstly, the proposed method was evaluated only on the NLST dataset, which contains approximately a thousand cancer cases. The limited sample size may have an effect on the model performance and further validation on other data sets is necessary. Secondly, due to the large size of each CT image, we adopted a pre-trained nodule detection model to extract the top five regions containing a nodule as input. While this reduced the computational burden, it led to the loss of some image information, and the accuracy of the nodule detection model may affect the results as well. Thirdly, our method only used medical images as input, and future studies can consider incorporating demographic and/or clinical information.

%\bibliographystyle{unsrt} % Bibliography style that orders references by appearance
\bibliography{references} % Bibliography file

\vfill

\end{document}